\documentclass[sigconf,nonacm]{acmart}

 \acmConference[CHI '23]{CHI '23: ACM CHI Conference on Human Factors in Computing Systems}{April 23--28, 2023}{Hamburg, Germany}
 \acmBooktitle{CHI '23: ACM CHI Conference on Human Factors in Computing Systems,  April 23--28, 2023, Hamburg, Germany}
\begin{document}

\title{Whose Text Is It Anyway? Exploring BigCode, Intellectual Property, and Ethics}

\author{Madiha Zahrah Choksi}
\email{mc2376@cornell.edu}
\orcid{0009-0008-4752-7164}

\affiliation{
  \institution{Cornell Tech}
  \streetaddress{2 West Loop Road}
  \city{New York}
  \state{New York}
  \country{USA}
  \postcode{10044}
}

\author{David Goedicke}
\email{dg536@cornell.edu}
\orcid{0000-0002-4837-893X}

\affiliation{
  \institution{Cornell Tech}
  \streetaddress{2 West Loop Road}
  \city{New York}
  \state{New York}
  \country{USA}
  \postcode{10044}
}

\renewcommand{\authors}{Choksi and Goedicke}

\begin{abstract}
 Intelligent or generative writing tools rely on large language models that recognize, summarize, translate, and predict content. This position paper probes the copyright interests of open data sets used to train large language models (LLMs). Our paper asks, how do LLMs trained on open data sets circumvent the copyright interests of the used data? We start by defining software copyright and tracing its history. We rely on GitHub Copilot as a modern case study challenging software copyright. Our conclusion outlines obstacles that generative writing assistants create for copyright, and offers a practical road map for copyright analysis for developers, software law experts, and general users to consider in the context of intelligent LLM-powered writing tools.
\end{abstract}

\ccsdesc[500]{Computer systems organization~Embedded systems}
\ccsdesc[300]{Computer systems organization~Redundancy}
\ccsdesc{Computer systems organization~Robotics}
\ccsdesc[100]{Networks~Network reliability}

\keywords{open-source, data, large language models, artificial intelligence, intellectual property, copyright, ethics}


\maketitle

\section{Software Copyright}
As a branch of intellectual property (IP) law, copyright protects unique expressions \cite{grimmelmann2004regulation}. Software, as a form of writing fixed in a tangible medium (source code), is protected under copyright law. As software and the prevalence of openly available code online scaled, two licensing models emerged to protect source code: open or free software licensing and closed or proprietary licensing. In the former free software model, source code is openly available to the public. In a general sense, anyone is welcome to use, share, or modify source code. In the latter model, source code is proprietary and protected from public access \cite{weber2005success}. This position paper is interested in probing the copyright interests of large language models that rely on open-source code at scale to produce intelligent writing tools. How do large language models trained on open-source data sets circumvent copyright interests? Does the doctrinal history of software licensing permit and protect this novel phenomenon?

\section{Tracing a Brief History of Software Copyright}

In 1978, the National Commission of New Technological Uses of Copyrighted Works (CONTU) released a lengthy report outlining how the Copyright Act of 1976 should treat software copyright\cite{samuelson1984contu}. CONTU was interested in computers and copyright: how can copyright extend to the creation of new works with computer assistance? What about the use of copyrighted works in conjunction with computers (i.e., copyrighted works placed into computers, the creation and use of automated databases, and the intellectual property of computer programs themselves)? In their final report, CONTU underscored protecting the incentives of code producers. CONTU argued that it takes more effort to create original creative works than to copy them, so the interests of the authors prevail \cite{samuelson1984contu}. CONTU reasoned that copyright law could easily expand to include new technologies and their creative capabilities, as copyright's underlying structures and goals are already laid out to protect owners and their creations. 

However, responses from developers within the open software community argued that treating software as the sole protected property of developers produces bad software \cite{moglenanarchism}. These communities maintained that source code is a non-literary form of composition, that code communicates to other programmers as it communicates to other readers, and that software is a product of collaborative institutions and relationships. Along these lines, the open-source community firmly believes software is for sharing. 

\section{Circling Back to Free-Software}
In contrast to the CONTU model of software copyright, the free and open-source software community relies on its own established framework for licensing. The open-source licensing model (encompassing copyleft and permissive licensing) protects source code through the use of licenses that specify the terms and conditions under which the software can be used, modified, and distributed. These licenses are designed to ensure that the software remains free and open-source and that anyone who uses the software is able to do so without restrictions or fees.

There are over eighty open-source licenses, and the most commonly applied include the GNU General Public License (GPL), the MIT License, and the Apache License. While each license has its own specific language, they share the common goal of protecting the freedom of the software.

Further, free software licenses require that the software be distributed with its source code and that any modifications or derivative works be distributed under the same license. Additionally, some licenses may require that any changes to the software be made available to the public or that any software that incorporates the free software also be made available under a compatible license.

Adopting a specific open-source license, modifying it, or writing an entirely new one is a process that signals community agreement or consensus around sharing and openness. Simply put, the community and the project are inextricably linked to the license that enacts and reinforces the given community's commitment to sharing, transparency, and openness. 

\section{Copilot: A Case Study of Writing Assistants}
In 2021, GitHub and OpenAI launched Copilot, a closed commercial “AI-pair programmer" plug-in that supports code composition and auto-completion \cite{CopilotIntro}. Copilot markets itself as a tool that helps programmers write better and more efficient code. From inserting console.logs() to debug code, identifying contextual patterns and auto-suggesting functions, and generating entire code blocks from comments, Copilot prides itself on supporting code generation for the GitHub community \cite{barke2022grounded}. The availability of open-source code on GitHub powers Copilot by ingesting incalculable code samples hosted on the platform released under various copyleft licenses. That being said, there is nothing philosophically iniquitous about employing open-source code or software for commercial purposes, and this ideological commitment (held by the open-source community) has been enshrined in numerous legal debates, including Google v. Oracle \cite{gratz2017platforms}. However, as a closed tool, Copilot treats software as property, making essential aspects of copyright law, such as attribution, impossible. 

\section{Conundrums}
However convenient Copilot's writing assistant may be, the smart writing assistant creates a serious copyright conundrum: a code recommendation system that generates copyright-infringing works. Copilot is trained on open-source code from repositories on GitHub, released under numerous open-source licenses without any attribution. According to GitHub, the text of the GPL was contained in the training data 700,000 times. Under U.S. copyright law, there can be no secondary copyright liability without primary infringement \cite{grimmelmann2015there}. This means that Copilot is not itself liable for copyright infringement until the tool is utilized to produce an output. Therefore, as a code-writing assistant, Copilot relies on copied code, creating a significant risk to its users.

Ongoing litigation will determine the legality of laundering open-source code to power an AI-writing tool with a paid subscription model. However, regardless of legality, a number of developers and projects have left GitHub, citing that the platform is violating the spirit of the open community, copyleft licensing, and, most notably, their trust. One complaint, for example, asks why the models treat open-source code as a free commodity. Why weren't these models trained on the proprietary code of GitHub and its owner Microsoft \cite{Kuhn_2022} if there are no copyright concerns with the system? 

Without thorough copyright analyses, there is also a heightened risk of normative consequences: a perceived precedence that entities training and distributing tools built using open-source code face no liability risks.

\section{Areas for Copyright Analysis}
We provide the following road-map to outline what a detailed copyright analysis of generative writing tools built on open-source code should carefully consider: 

\begin{description}
\item[1. License review:] GitHub houses many different open-source repositories, yet it is unclear how the applied licenses permit the use of the code for a system like Copilot. A thorough license review will reveal which repositories permit copying of code (MIT), which do not without attribution (GPL), which are ambiguous, and which have been modified by developers with further specifications. This review, against the license that Microsoft and Open AI release with Copilot, will signal if and how Copilot assumes copyright interest over source code that is already protected. 

\item[2. Fair use analysis:] Fair use is strictly defined and permits using copyrighted material \textit{without} optioning the copyright. Under fair use, any person may use certain amounts of copyrighted material without permission from the copyright owner. Fair use thrives in numerous domains, but most prominently in educational uses. Within this context, copyrighted works are permitted for limited use for educational purposes. Analyzing the goals and purpose of the writing tool and the training data it relies upon may demonstrate that it is protected under the fair use doctrine. In other words, there may be uses of the writing assistant that fit under fair use and others that would make the tool lose fair use status.

\item[3. Attribution conflicts:] The explicit purpose of attribution is to provide reciprocal incentives to open-source code authors or developers. If a work is protected by copyright, the author must be attributed when directly quoting portions of their work. As soon as original copyrightable data is extracted from a repository, this feedback channel is broken. The absence of attribution can significantly impact open communities, as it stymies motivation for knowledge production and dissemination.

\item[4. Data privacy issues:] First, inherent to large language models (particularly those that produce closed applications) is the inability to check and verify that certain sensitive data is not stored. Models could be trained on data that contains personally identifiable data or data that can be de-anonymized. This data can be leaked to users through prompt-based LLM interfaces. Second, open repositories can be scraped and filtered, and computational inference methodologies can be used to derive patterns and infer conclusions. These include (but are not limited to) provenance and data subjects that may have been omitted from the open data set. Exploring data privacy issues is vital to understanding risks and the copyright interests of the developers who publish these data sets and those whom the data concerns.

\end{description}
Copyright law has extended to encompass written code and software programs and applications. However, the rise of intelligent AI-based writing tools trained on vast amounts of open-source licensed code creates a modern dilemma for copyright. Our paper attempts to define this dilemma as one copyright experts have previously dealt with (CONTU). By starting with the broader history of software copyright and open-source licensing, we offer a framework for copyright analysis that addresses both the material used for training models and the material generated by the algorithm. Our analysis can be a starting point for developers creating and interfacing with LLM-based tools in the absence of clear legal guidelines and risks for intelligent LLM-powered tools. 

\balance

\bibliographystyle{ACM-Reference-Format}
\bibliography{bibliography}

\end{document}